\documentclass[twocolumn]{article}

\usepackage{authblk}

\usepackage{footnote}
\usepackage{graphicx}
\usepackage{paralist}
\usepackage{url}
\usepackage[toc,page]{appendix}
\usepackage{algpseudocode}
\usepackage{algorithm}

\usepackage{color}

\begin{document}

\title{An iterative technique to identify browser fingerprinting scripts}

\author[1,2]{Antonin Durey}
\author[3,1,2]{Pierre Laperdrix}
\author[1,2]{Walter Rudametkin}
\author[1,2,4]{Romain Rouvoy}

\affil[1]{University of Lille}
\affil[2]{Inria}
\affil[3]{CNRS}
\affil[4]{IUF}
\affil[ ]{\textit {\{firstname.lastname\}@univ-lille.fr}}

\date{}

\maketitle

\noindent\textbf{Abstract---}\emph{Browser fingerprinting} is a stateless identification technique based on browser properties.
Together, they form an identifier that can be collected without users' notice and has been studied to be unique and stable.
As this technique relies on browser properties that serve legitimate purposes, the detection of this technique is challenging.
While several studies propose classification techniques, none of these are publicly available, making them difficult to reproduce.
This paper proposes a new browser fingerprinting detection technique.
Based on an incremental process, it relies on both automatic and manual decisions to be both reliable and fast.
The automatic step matches API calls similarities between scripts while the manual step is required to classify a script with different calls.
We publicly share our algorithm and implementation to improve the general knowledge on the subject.


\vspace{-10pt}

\section{Introduction}

As cookies and stateful identification techniques become more and more restricted in browsers~\cite{firefox63cookies, safaricookies}, new techniques have emerged to identify users across the web.
Among them, \emph{browser fingerprinting} is an identification technique based on the collection of browser properties.
It is permissionless, unoticeable for the users and has been studied to be unique~\cite{eckersley10, laperdrix16, gomezboix18} and stable~\cite{vastel20}.
While identifying cookie uses by scripts is easy, browser fingerprinting relies on dozens of JavaScript APIs that have legitimate usages when used separately.
Therefore, the identification of this technique is difficult.

We propose here a incremental technique to identify browser fingerprinting by classifying an input dataset made of scripts.
We defined 3 labels to be assigned to scripts: \emph{fingerprinter}---or suspect---for scripts that we consider doing fingerprinting, \emph{non-fingerprinter}---or clean---for scripts not collecting a fingerprint, and \emph{unknown} for scripts we do not have enough information about to make a decision.
We used the browser fingerprinting scripts identified by \textsc{Disconnect} to provide a ground truth and initiate our algorithm with a list of \emph{fingerprinters}~\cite{disconnecttrackingprotection}.
We then computed a similarity score between scripts from the input dataset and \emph{fingerprinters} based on API calls that are known to be used for fingerprinting.
Using the similarity score, we could automatically classify the script as \emph{fingerprinter} or \emph{non-fingerprinter}.
If the automatic algorithm could not be able to label a script, we propose a manual phase to label the script.
At the end of the execution, each script of the input dataset is labeled \emph{fingerprinter}, \emph{non-fingerprinter} or \emph{unknown}.

The rest of this report is organized as follow.
We present the state of the art in Section~\ref{sec:background}.
We describe our technique in Section~\ref{sec:classification}.
We discuss our technique in Section~\ref{sec:discussion} and conclude in Section~\ref{sec:conclusion}.

\section{Background \& related work}\label{sec:background}

\subsection{Browser fingerprinting}

Browser fingerprinting is a technique to identify a user based on its hardware characteristics and browser configurations.
It combines properties to build an identifier~\cite{laperdrix19} such as the HTTP headers~\cite{eckersley10}, navigator and screen JavaScript properties~\cite{eckersley10, laperdrix16}, canvas~\cite{mowery12} and WebGL rendering~\cite{cao17}.
Studied by Eckersley~\cite{eckersley10} and Laperdrix~\emph{et al.}~\cite{laperdrix16}, the technique can lead to a unique identifier.
Moreover, Vastel~\emph{et al.}~\cite{vastel18stalker} showed it is possible to rely on some stable properties to obtain a stable fingerprint and track users accross the web for a long period of time.
The identification property of a fingerprint can lead either to a tracking technique for malicious purposes~\cite{englehardt16} or to enhance web security by improving authentication~\cite{alaca16} or detecting bots~\cite{vastel20}.


\subsection{Classification of fingerprinters}
Since its inception in 2010, browser fingerprinting also triggered many research contributions to identify the characteristics of scripts implementing browser fingerprinting techniques.
These contributions mostly lead to the publication of lists of URLs or common patterns.
For example, several browsers or extensions use Disconnect\footnote{\url{https://github.com/disconnectme/disconnect-tracking-protection}} or Easy Privacy\footnote{\url{https://github.com/easylist/easylist/tree/master/easyprivacy}} to protect users against fingerprinters.
Some studies focused on classifying browser fingerprinting scripts (or \emph{fingerprinters})~\cite{acar14, englehardt16}.
Unfortunately, these studies only focus on a reduced set of attributes and are not relevant when studying a real-world dataset adopting advanced browser fingerprinting techniques.
Bird~\emph{et~al.}~\cite{bird20} used API call similarities to detect fingerprinters and train a machine learning model.
However, their ground truth consists of specific attributes or keyword lists that cannot be considered reliable when labeling a real-world dataset. 
More recently, Iqbal~\emph{et~al.}~\cite{iqbal20} and Rizzo~\emph{et~al.}~\cite{rizzo21} proposed interesting approaches to classify fingerprinting scripts.
Both studies combined static and dynamic analysis to build a machine learning classifier.
They both relied on a manual analysis to improve the classifier over time.
Another more recent work aims at building signatures based on the script's behaviour~\cite{chen20}.
While we believe their approach is the most advanced one to distinguish malicious from benign scripts, it has not been used to classify browser fingerprinting scripts.
moreover, the code of all of these approaches is not publicly available, making them difficult to reproduce and reuse.

\section{Classification of Fingerprinters: an incremental method}\label{sec:classification}

The efficiency of browser fingerprinting relies on the design and implementation of a wide combination of stateless attributes, which can uniquely identify a user that visits a web page.
However, this combination of fingerprinting attributes is not formally defined and constantly changes due to the evolution of JavaScript APIs.
The list of attributes we collected covers all the attributes currently reported in the literature, but this list might evolve as some API might become deprecated or being removed by browsers vendors.
Oppositely, new APIs introduced in browsers might be reported by the community as being usable in a fingerprinting context.
This makes browser fingerprinting challenging to detect at large.
As we mentioned in Section~\ref{sec:background}, some techniques exist~\cite{iqbal20, rizzo21} to classify fingerprinting scripts, but their implementation is not publicly available, leading to a technique hardly reusable to label a real-world dataset using all the attributes a browser fingerprint contains.
This section, therefore, applies a supervised classification technique that leverages a ground truth of known browser fingerprinting scripts to label a collection of \emph{unknown} scripts as \emph{fingerprinter} or \emph{non-fingerprinter}, based on the combination of APIs and parameters they access.
We use similar techniques and methodology reported in the litterature~\cite{iqbal20, rizzo21}, provide our classification algorithm in Appendix~\ref{appendix:algorithm} and the implementation online.\footnote{\url{https://github.com/antonin-durey/fingerprinting-classification}}

\begin{figure*}[th]
    \centering
    \includegraphics[width=\textwidth]{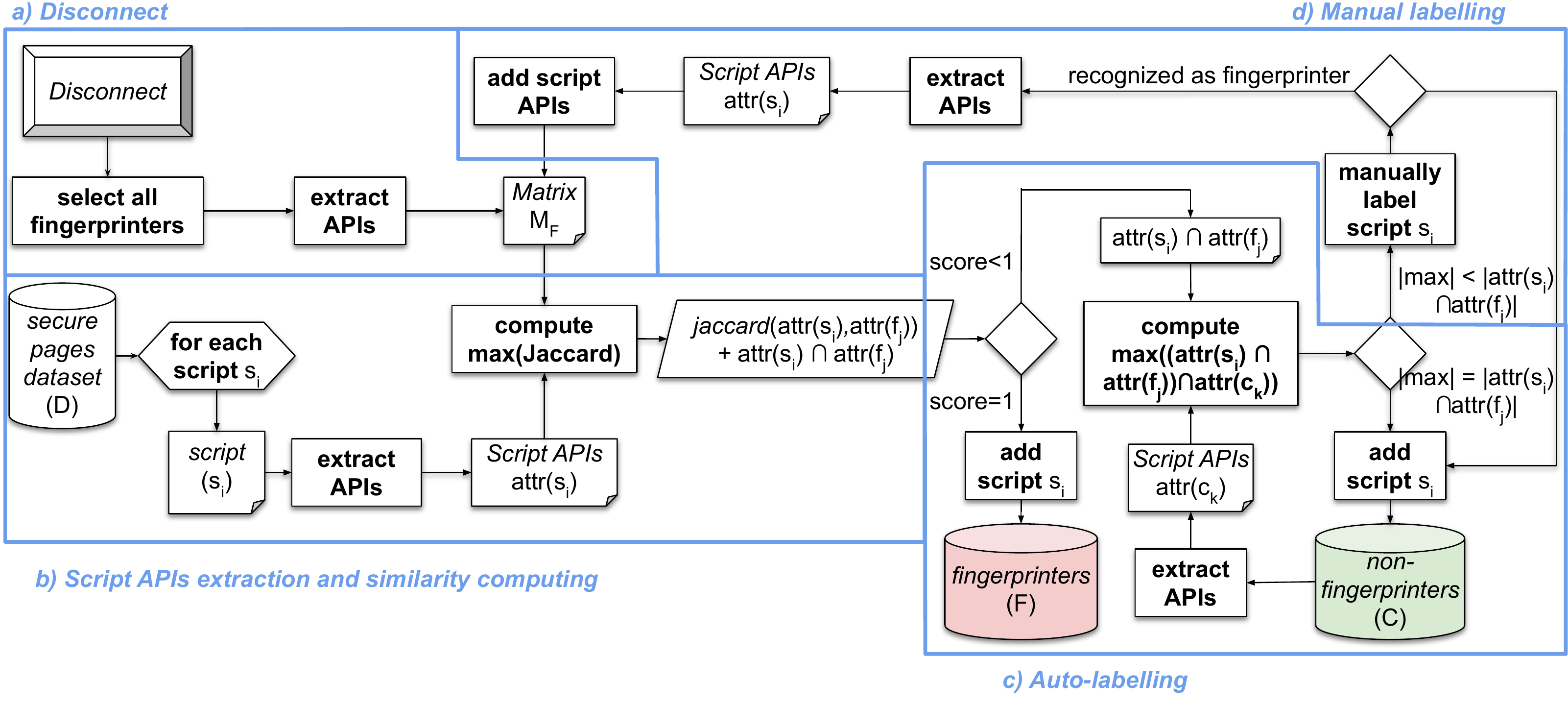}
	\vspace{-6pt}
    \caption{Flow chart representing our incremental script classification algorithm.}
    \label{fig:flowchart}
\end{figure*}

Given the lack of a formal model to identify browser fingerprinting scripts, we propose to adopt an incremental classification process.
In particular, we leverage a list of scripts classified as \emph{fingerprinter} by \textsc{Disconnect} to explore similarities in the accessed APIs.
Therefore, we compute the similarity score (Jaccard index) to automatically classify our scripts.
If a script cannot be labeled automatically, we go through a manual analysis to classify it.
We continuously update the similarity scores whenever a script is classified as \emph{fingerprinter} or \emph{non-fingerprinter}.
This way, we reduce the number of iterations required to label all the scripts included in our dataset by exploiting script similarities.
Figure~\ref{fig:flowchart} provides an overview of our incremental script classification approach.

\subsection{Learning fingerprinting attributes from \textsc{Disconnect}}
We bootstrap our approach with the \textsc{Disconnect} project that provides a list of fingerprinters whose behavior has been analyzed by experts~\cite{disconnecttrackingprotection}.
\textsc{Disconnect} has the advantage of being public, constantly updated, and recognized by the community to be reliable~\cite{bird20, iqbal20, rizzo21}.

Our first step consists of extracting all the JavaScript attributes $attr(f_i)$ accessed by each of the  fingerprinters $f_i \in F$ reported by \textsc{Disconnect} to identify the discriminating features of known fingerprinters.
We bypass minification and obfuscation techniques by instrumenting and monitoring the runtime behavior of each script loaded in an empty web page.
Each signature of a fingerprinting attribute is structured as a 3-tuple ${\langle}name,args,N{\rangle}$, where $name$ is the name of the accessed API, $args$ is he list of parameters used to retrieve this attribute (empty if none) and $N$ is the number of times that the pair ${\langle}name,args{\rangle}$ has been observed along with the execution of script $f_i$.
This results in a matrix of attributes that characterize the fingerprinters listed by \textsc{Disconnect}, $M_F$.

\subsection{Computing the fingerprinter similarity for unknown scripts}
We start an incremental classification process that takes as input the classification matrix of known fingerprinters ($M_F$) and the list of scripts ($S$) from our dataset of secure web pages.
For each script $s_i \in S$:
\begin{compactenum}
    \item we extract the set of attributes for each script ($attr(s_i)$),
    \item we compute the similarity score (Jaccard index) between $attr(s_i)$ and the attributes $attr(f_j)$ of each known fingerprinter $f_j \in F$, and
    \item we keep the tuple with the maximum similarity score and its corresponding intersection\\
    ${\langle}$ $s_i$ $,$ $jaccard(attr(s_i),attr(f_j))$ $,$ $attr(s_i){\cap}attr(f_j))$ ${\rangle}$.    
\end{compactenum}
We order the scripts from $S$ by decreasing similarity score to iteratively find the closest known fingerprinters to this script.

\subsection{Auto-labelling}
If the maximum similarity score of $s_i$ equals 1, $s_i$ implements a browser fingerprinting feature and we automatically label it as such.
However, if the score is less than 1, we need to compare to all of the non-fingerprinting scripts already labeled to take a decision.
For each \emph{non-fingerprinter} $c_k \in C$, we calculate a new intersection (that reuses the previous one we saved), specifically $attr(c_k)\,\,{\cap}\,\,\,(attr(s_i){\cap}attr(f_j))$, and we keep the result that maximizes the size of the new intersection.
If the attributes we obtain from our intersection with fingerprinters are the same attributes we obtain from our intersection with non-fingerprinters, \textit{i.e.,} the intersections are equal, the script's fingerprinting attributes are not discriminating enough to be considered a fingerprinter, so we label the script $s_i$ as a \emph{non-fingerprinter} script.
However, if the intersections differ, the fingerprinting attributes $attr(s_i)$ are new in the classification, and our algorithm cannot automatically label the script $s_i$.

\subsection{Manual labeling}
As pointed out by previous studies~\cite{bird20, iqbal20, rizzo21}, manual labeling is necessary when the automatic tools used to classify are unable to decide the label of a script.
If our algorithm cannot automatically label the script, we manually analyze the code and label it as either \emph{fingerprinter}, \emph{non-fingerprinter}, or \emph{unknown}.
To do so, we use the following criteria:
\begin{compactitem}
    \item the script is blocked by \textsc{EasyList} or \textsc{EasyPrivacy},\footnote{\url{https://github.com/easylist/easylist}} 
    \item the script contains obvious keywords that reveal its goal (\textit{e.g.,} \textit{fingerprinting}, \textit{deviceFingerprint}),
    \item the attribute values are forwarded to a remote server, as it can be used to compute similarities with previously saved fingerprints, or
    \item the privacy policy, when available, of the company owning the script mentions \emph{fingerprinting}, \emph{stateless identification technique}, \emph{device} or \emph{browser identification}.
\end{compactitem}

As soon as the manual evaluation matches 2 of the above criteria, we label the script as \emph{fingerprinter}.
Whenever we label a \emph{fingerprinter}, the fingerprinting attributes of $s_i$ are added to the detection matrix $M_F$.
Each time we rerun the classification process in a new iteration and stop when all the scripts of our dataset have been labeled.
The classification algorithm can be found in appendix~\ref{appendix:algorithm}.

\section{Threat to validity}\label{sec:discussion}

Script classification only relies on what is observed on a client-side.
The backend treatment applied to the collected data cannot be analyzed.
Thus, it is impossible to understand with certitude the usage of the browser fingerprinting technique.
While classification techniques tend to estimate a script goal, they suffer several threats:

\noindent\textbf{Attributes.}
It is not possible to monitor accesses to HTTP headers as they are systematically sent with every request, whether or not the website uses them.
They can be used by themselves or in aggregation with other fingerprinting techniques.
Even if the HTTP headers alone cannot uniquely identify users~\cite{gomezboix18}, they can still be used to add entropy.

\noindent\textbf{The ground truth.}
Our classification algorithm leverages the distances computed between our dataset and the ground truth we chose.
While \textsc{Disconnect} is considered as reliable by the litterature~\cite{iqbal20,bird20,rizzo21}, we cannot guarantee the results given by our algorithm would have been the same if we decided to use a different ground truth.

\noindent\textbf{Manual labelling.}
Concerning manual labeling, other experts in the domain could have different opinions about some scripts, leading to different labels in ambiguous cases.
For example, the collection of \texttt{navigator} and \texttt{screen} properties can help a website adapts itself to the device and browser of a user.
Some users could consider this as fingerprinting while other could estimate this data collection legitimate for the website.
Thus, we expect several users of our technique to obtain different results according to their conception of browser fingerprinting.

\noindent\textbf{Similar scripts.}
Our algorithm labels scripts based on the similarity between the attributes collected by a script.
Because we based our approach on fingerprinting attributes, we assumed 2 scripts collecting the same attributes should have the same label.
This restriction might have only minor consequences for scripts that collect dozens of attributes and show little ambiguity, however, scripts with a limited number of attributes might use the same attributes for different purposes.

\section{Conclusion}~\label{sec:conclusion}

In this paper, we proposed a new methodology to classify browser fingerprinting script.
Based on an iterative approach, the technique uses the advantage of both automatic and manual labelling to label a dataset of scripts.
This technique can be used to classify a wild dataset from scripts obtained of real websites.
We made our algorithm and implementation available.\footnote{\url{https://github.com/antonin-durey/fingerprinting-classification}}

\bibliographystyle{plain}
\bibliography{references}

\appendix

\section{Classification Algorithm}\label{appendix:algorithm}

\begin{algorithm*}
    \caption{Script classification algorithm}
    \begin{algorithmic}[1]
    \Function{Classify}{$ disconnect, dataset $}
		\State $ fingerprinters \gets \Call{BuildFrom}{disconnect}$
        \State $ suspectScripts \gets \left\langle{}\right\rangle $
        \State $ cleanScripts \gets \left\langle{}\right\rangle $
        \State $ unknownScripts \gets \left\langle{}\right\rangle $
        \State $ hasEnded \gets False $
        \While {$ !hasEnded $}
	        \State $ needToRecompute \gets False $
            \State $ scriptsScoreAndIntersection \gets \Call{ComputeScoreAndIntersection}{fingerprinters, dataset} $
            \State $ \Call{SortByScoreDesc}{scriptsScoreAndIntersection } $
            \State $ i \gets 0 $
            \While{$ !needToRecompute And i < \Call{Length}{scriptsScoreAndIntersection}$}
            	\State $script, score, fingerprintersIntersection \gets scriptsScoreAndIntersection[i] $
            	\If {$ score = 1 $}
					\State $ suspectScripts \gets suspectScripts \cup \left\langle{script}\right\rangle $	
				\Else 
					\State $ cleanIntersection \gets \Call{ComputeBiggestIntersection}{cleanScripts, intersection}$	     
					\If {$\Call{Equals}{fingerprintersIntersection, cleanIntersection}$}
                        \State $ cleanScripts \gets cleanScripts \cup \left\langle{script}\right\rangle $	
	                \Else
                    	\State $ label \gets \Call{ManuallyLabelScript}{script} $
                    	\If{$\Call{IsFP}{label} $}
                        	\State $ fingerprinters\gets fingerprinters \cup \left\langle{script} \right\rangle $
					        \State $ needToRecompute \gets True $
                    	\ElsIf {$ \Call{IsNonFP}{label} $}
                        	\State $ cleanScripts  \gets cleanScripts \cup \left\langle{script} \right\rangle $
	        				\State $ needToRecompute \gets True $
                    	\ElsIf {$ \Call{IsUnknown}{label} $}
                        	\State $ unknownScripts \gets unknownScripts \cup \left\langle{script} \right\rangle $
						\EndIf
                    \EndIf
                \EndIf
                \State $i \gets i + 1$
            \EndWhile
            \If{$i == \Call{Length}{scriptsScoreAndIntersection}$}
		        \State $ hasEnded \gets True $
            \EndIf
        \EndWhile
        \State \Return $ suspectScripts, cleanScripts, unknownScripts $
    \EndFunction
    \end{algorithmic}    
\end{algorithm*}

\end{document}